\documentstyle[aps,pre,epsf]{revtex}

\newcommand{\figurewidth}{13cm}

\begin{document}

\draft

\title{Critical properties of the three-dimensional equivalent-neighbor model\\
       and crossover scaling in finite systems}
\author{Erik Luijten\thanks{Electronic address:
        erik.luijten@uni-mainz.de}}
\address{Department of Physics, Delft University of Technology, P.O. Box
         5046, 2600 GA Delft, The Netherlands}
\address{Max-Planck-Institut f\"ur Polymerforschung, Postfach 3148,
         D-55021 Mainz, Germany}
\address{Institut f\"ur Physik, WA 331, Johannes Gutenberg-Universit\"at,
         D-55099 Mainz, Germany\thanks{Present address.}}

\date{November 23, 1998; corrected January 29, 1998}

\maketitle

\begin{abstract}
  Accurate numerical results are presented for the three-dimensional
  equivalent-neighbor model on a cubic lattice, for twelve different
  interaction ranges (coordination number between 18 and~250). These results
  allow the determination of the range dependences of the critical temperature
  and various critical amplitudes, which are compared to renormalization-group
  predictions.  In addition, the analysis yields an estimate for the
  interaction range at which the leading corrections to scaling vanish for the
  spin-$\frac{1}{2}$ model and confirms earlier conclusions that the leading
  Wegner correction must be negative for the three-dimensional
  (nearest-neighbor) Ising model.  By complementing these results with Monte
  Carlo data for systems with coordination numbers as large as~$52514$, the
  full finite-size crossover curves between classical and Ising-like behavior
  are obtained as a function of a generalized Ginzburg parameter. Also the
  crossover function for the effective magnetic exponent is determined.
\end{abstract}

\pacs{64.60.Fr, 75.40.Cx, 75.10.Hk, 05.70.Fh}


\section{Introduction}
\label{sec:intro}
Over the past decades, several techniques have been applied to investigate how
the critical behavior of systems depends on the range of the interactions.
Before the general acceptance of the concept of universality, it was not at all
clear that the critical properties of all systems with a one-component order
parameter and ferromagnetic (i.e., attractive) interactions with a {\em
finite\/} range are described by the Ising universality class. Since it was
realized that most interactions in nature are not necessarily restricted to the
nearest neighbors, one thus tried to determine the properties of models with a
larger coordination number~$q$. Another motivation, which plays a more
important r\^ole in the present work, is the fact that in the limit of infinite
interaction range one recovers the classical or mean-field model. Since the
latter model can be solved analytically whereas no exact solution has been
found for three-dimensional systems with a finite interaction range~$R$, it is
of interest to see how the crossover takes place from finite to infinite~$R$.
A natural choice for the examination of this crossover is the so-called
``equivalent-neighbor'' model, introduced by Domb and Dalton~\cite{domb66}.  In
this generalization of the Ising model, each spin interacts equally strongly
with all its neighbors within a certain distance, whereas all remaining
interactions are equal to zero. In Ref.~\cite{domb66}, series expansions have
been used to investigate two- and three-dimensional systems with interactions
extending up to the third shell. On a simple cubic lattice this corresponds to
26~neighbors and on a face-centered cubic lattice even to 42~neighbors. While a
general trend toward the mean-field properties, especially for the critical
temperature, is clearly visible from these results, several problems emerge.
First, with increasing interaction range, increasingly longer series are
required to achieve a certain degree of convergence. Secondly, it appears that
the maximum coordination numbers examined by this method are not large enough
to observe the asymptotic deviations from the mean-field
behavior~\cite{thouless}. Although Ref.~\cite{domb66} was published over
30~years ago, it appears that, especially in three dimensions, no substantial
progress toward larger coordination numbers has been pursued. This is probably
caused by the fact that also other techniques are plagued by serious
difficulties upon increase of the interaction range. For example, Monte
Carlo~(MC) methods in general suffer from a serious decrease in efficiency if
the number of interactions increases. Mon and Binder~\cite{mon} studied
two-dimensional~(2D) spin systems with a maximum coordination number $q=80$,
compared to $q=12$ and $q=18$ for quadratic and triangular lattices,
respectively, in Ref.~\cite{domb66}. Furthermore, they derived the
$R$~dependence of critical amplitudes from scaling considerations.  However, it
still proved difficult to reach the asymptotic regime where the predictions
were expected to hold. In a subsequent paper~\cite{medran}, Luijten {\em et
al.}\ confirmed the predictions of Ref.~\cite{mon} from a
renormalization-group~(RG) analysis and revealed the existence of a logarithmic
$R$~dependence in the shift of the critical temperature. Thanks to the advent
of a dedicated MC algorithm for long-range interactions~\cite{ijmpc}, systems
with large coordination numbers could be simulated without loss of efficiency.
Thus, in the same paper the critical properties were determined for quadratic
systems with coordination numbers up to $q=436$. It was explicitly verified
that all examined systems belong to the 2D~Ising universality class and the
predicted $R$~dependence of the critical amplitudes could indeed be observed,
as well as the approach of the critical temperature toward its mean-field
value. It is the purpose of the present work to extend this analysis to
three-dimensional~(3D) systems. Apart from the possibility to verify the
predicted range dependences in three dimensions, a precise knowledge of the
critical properties of spin models with an extended range of interaction also
serves a further purpose.  Namely, it allows the study of two forms of
crossover in these systems. {\em Finite-size crossover\/} only pertains to
finite systems at the critical temperature and denotes the transition from the
classical regime where the interaction range is at least of the order of (some
power of) the system size to the nonclassical (Ising) regime where the system
size is much larger than the interaction range.  {\em Thermal crossover}, on
the other hand, occurs when the temperature is moved away from its critical
value.  The interplay between the range~$R$ of the interactions and the
decreasing correlation length~$\xi$ determines the location of the crossover to
classical critical behavior.  If $R$ is small, the temperature distance to the
critical temperature $T_c$ must be made rather large before $\xi$ and $R$ are
of the same order of magnitude. In such systems, no crossover to
mean-field-like critical behavior can be seen, because one has already left the
critical region.  However, for $R$ large, it is well possible to observe both
Ising-like and classical critical behavior. This dependence on both $t \equiv
(T-T_c)/T_c$ and $R$ is expressed by the Ginzburg criterion~\cite{ginzburg}.
Both variants of crossover have been studied for 2D systems in
Refs.~\cite{chicross,cross}, which showed that accurate information on
crossover scaling functions can be obtained by numerical techniques. In the
light of a comparison to experimental results on the one hand and theoretical
calculations of crossover scaling functions on the other hand, it is extremely
relevant to investigate the 3D case as well.  Here, I present the results of MC
simulations of systems with interactions up to a distance of $\sqrt{14}$
lattice units (thirteenth shell), which corresponds to 250~equivalent
neighbors.  Although larger interaction ranges do not diminish the efficiency
of the MC algorithm, an accurate determination of the critical properties for
larger~$R$ is hampered by a different effect. Indeed, such a determination is
only possible in the Ising limit, which implies that the {\em smallest\/}
linear system sizes must be of the order of $L_{\rm min}={\cal O}
(R^{4/(4-d)})$~\cite{medran}, where $d$ indicates the dimensionality. Thus, for
$d=3$ the smallest allowable systems contain of the order of $R^{12}$ spins and
one can only hope that this relation exhibits a prefactor considerably smaller
than unity.

The results of the MC simulations are then used to determine the finite-size
crossover functions for several quantities.  It should be noted that for a full
mapping of these function {\em very\/} large coordination numbers are required:
simulations have been carried out for~$q$ up to $52514$.  Yet, an independent
determination of the critical temperature of these systems is not required, but
can be obtained by extrapolation. It suffices thus to study modest ($L \leq
40)$ system sizes for these interaction ranges. The determination of thermal
crossover functions will be the subject of a future paper~\cite{cross3d}, as it
requires calculations which are actually complementary to those of the present
work (results for the susceptibility can be found in Ref.~\cite{chi3d}).
Indeed, for a determination of the critical properties by finite-size scaling
and for the mapping of the finite-size crossover functions, all data must lie
within the finite-size regime, whereas for thermal crossover scaling care must
be taken that the data lie outside this regime.

Two further questions that are addressed in this paper concern the corrections
to scaling. In the first place, the range dependence of the thermal finite-size
corrections is shown to be in very good agreement with the predictions of
Ref.~\cite{medran}. Secondly, the finite-size corrections due to the leading
irrelevant field are analyzed and the related variation of the $\phi^4$
coefficient in the Landau-Ginzburg-Wilson~(LGW) Hamiltonian is obtained. This
permits an estimation of the interaction range for which this coefficient
coincides with its fixed-point value and confirms that for the
three-dimensional nearest-neighbor Ising model it does not lie between the
Gaussian fixed point and the Ising fixed point.

The outline of this paper is as follows. In Sec.~\ref{sec:summary}, I briefly
summarize earlier predictions for the range dependence of critical amplitudes
and discuss the shift of the critical temperature as a function of interaction
range. Section~\ref{sec:mc} gives details of the Monte Carlo simulations.
Furthermore, the determination of the critical temperatures is discussed as
well as the analysis of the range dependence of corrections to scaling. The
variation of critical amplitudes as a function of interaction range is treated
in Sec.~\ref{sec:rangedep} and finite-size crossover curves are obtained in
Sec.~\ref{sec:cross}. I end with some concluding remarks in
Sec.~\ref{sec:concl}.

\section{Summary of renormalization-group predictions}
\label{sec:summary}
In the absence of an external field, the equivalent-neighbor or medium-range
model is defined by the following Hamiltonian,
\begin{equation}
 {\cal H}/k_B T = -\sum_{\langle ij\rangle} K({\bf r}_i-{\bf r}_j) s_i s_j \;,
\label{eq:hamil}
\end{equation}
where $s = \pm 1$, the sum runs over all spin pairs and the spin-spin coupling
is defined as $K({\bf r}) = J > 0$ for $|{\bf r}| \leq R_m$ and $K({\bf r}) =
0$ for $|{\bf r}| > R_m$.  I first summarize the findings of Ref.~\cite{medran}
for the $R$ dependence of critical properties, as obtained by an RG analysis.
Although at first sight this approach is not very different from a simple
scaling analysis, it offers several advantages. The formulation in terms of two
competing fixed points provides a clear insight into the crossover mechanism:
for $R$ large the coefficient of the $\phi^4$ term in the LGW Hamiltonian is
suppressed with respect to the quadratic term in this expression. Thus, the
renormalization trajectory passes close to the Gaussian fixed point and the
critical amplitudes pick up a specific $R$ dependence which is determined by
the flow near this fixed point.  For any finite $R$, the system will still flow
to the neighborhood of the nontrivial (Ising) fixed point (cf.\ Fig.~1 in
Ref.~\cite{medran}). However, the $R$ dependence reveals some aspects of the
Gaussian fixed point which are not normally seen in Ising-like systems. For
example, near this fixed point the thermal exponent $y_t$ and the leading
irrelevant exponent $y_i$ assume the values 2 and $4-d$, respectively, which
coincide for $d=2$. Such a coincidence would lead to logarithmic factors in the
scaling functions, were it not that the Gaussian fixed point is unstable for
$d=2$. In contrast, the $R$ dependence of scaling functions indeed allows the
observation of such logarithms. The occurrence of these dependences is not
easily found from a scaling analysis.

For the magnetization density~$m$ and the magnetic susceptibility~$\chi$ the
following range dependences have been obtained,
\begin{eqnarray}
\label{eq:mag-rangedep}
m    &\propto& t^\beta R^{(2d\beta-d)/(4-d)}      \;, \\
\label{eq:chi-rangedep}
\chi &\propto& t^{-\gamma} R^{2d(1-\gamma)/(4-d)} \;,
\end{eqnarray}
where $\beta$ and $\gamma$ denote the standard Ising critical
exponents. Furthermore, the finite-size scaling behavior of these quantities
was derived as
\begin{eqnarray}
\label{eq:mag-fss}
m    &=& L^{y_h-d} R^{(3d-4y_h)/(4-d)}
         \hat{f}^{(1)}_s
         \left( tL^{y_t} R^{-2(2y_t-d)/(4-d)},
                \tilde{u}L^{y_i} R^{-4y_i/(4-d)},
                hL^{y_h} R^{(3d-4y_h)/(4-d)}
         \right) \;, \\
\chi &=& L^{2y_h-d} R^{2(3d-4y_h)/(4-d)}
         \hat{f}^{(2)}_s
         \left( tL^{y_t} R^{-2(2y_t-d)/(4-d)},
                \tilde{u}L^{y_i} R^{-4y_i/(4-d)},
                hL^{y_h} R^{(3d-4y_h)/(4-d)}
         \right) \;.
\label{eq:chi-fss}
\end{eqnarray}
Here, $\hat{f}^{(i)}_s$ denote universal scaling functions, $y_t$ and $y_i$ are
the thermal and leading irrelevant exponent introduced before, and $y_h$ is the
magnetic exponent. $\tilde{u}$ and $h$ are the irrelevant and the magnetic
scaling field, respectively.

Also the shift of the critical temperature with respect to its mean-field value
has been calculated in Ref.~\cite{medran}. However, this treatment left several
questions unanswered, which I will consider here in some more detail. A clear
understanding of the nature of this shift is of particular significance for the
crossover scaling, since one has to calculate the critical temperatures for
systems with large coordination numbers by means of extrapolation.  It was
derived that under a renormalization transformation the contribution of the
$\phi^4$ term to the quadratic term in the LGW Hamiltonian leads to a
range-dependent shift of the reduced temperature $t \equiv (T-T_c)/T_c$. For
$d=2$ it was found in Ref.~\cite{medran} that this shift has the form
\begin{equation}
 T_c - T_c^{\rm MF} = \frac{c_0 + c_1\ln R}{R^2} + \cdots \;.
\label{eq:shift_2d}
\end{equation}
where $c_0$ and $c_1$ are constants. This expression has also been confirmed
numerically, see Fig.~4 in Ref.~\cite{medran}. Interestingly, this result has
been recovered in Ref.~\cite{caracciolo}, where in addition it was found that
the constant~$c_1$ has a universal value~$-2/\pi \approx -0.6366$.  Indeed,
this agrees with the value $-0.624~(7)$ obtained from an analysis of the
available data for $25 \lesssim R^2 \lesssim 70$. (The somewhat lower value
$0.609$, corresponding to the coefficients quoted in Ref.~\cite{cross}, can be
explained from the influence of the data point at $R^2=16.2$.)  However, the
result for general $2<d<4$, a shift proportional $R^{-2d/(4-d)}$, clearly
contradicts the results obtained from systematic expansions in terms of the
inverse coordination number (but see the remarks at the end of this section).
Brout~\cite{brout} obtained to leading order a shift of the form $1/q \propto
1/R^d$. This result was recovered by Vaks {\em et al.\/}~\cite{vaks} and Dalton
and Domb~\cite{dalton66}. As indicated in Ref.~\cite{thesis}, such an
additional and actually dominant shift can also be obtained from the RG
analysis by allowing for a (spherically symmetric) lower-distance cutoff $a$ in
the spin-spin coupling $K({\bf r})$. In momentum space the coupling then takes
the form
\begin{equation}
  \tilde{K}({\bf k})
  = c \left(\frac{2\pi}{kR}\right)^{d/2} J_{d/2} (kR) -
  c\left(\frac{a}{R}\right)^d\left(\frac{2\pi}{ka}\right)^{d/2} J_{d/2} (ka)\;,
\end{equation}
where $c=JR^d$ and $J_\nu$ is a Bessel function of the first kind of order
$\nu$ (cf.\ Eq.~(A3) in Ref.~\cite{medran}). The second term in this expression
yields an additional contribution to the quadratic term in the LGW Hamiltonian,
which is precisely the $1/R^d$ shift obtained by Brout.  Furthermore, it
contributes to the $k$-dependent part of this term, which via the rescaling of
the field (see Ref.~\cite{medran}) leads to a $1/R^{d+2}$ shift. Note that,
upon expansion in powers of~$R$, a formulation in terms of the coordination
number~$q$ implies such a shift as well.  At even higher order, one finds (at
rational dimensionalities) additional $\ln R$ dependences, as was first
recognized by Thouless~\cite{thouless}.\footnote{This work only came to the
attention of the author after the publication of Ref.~\protect\cite{medran}.}
He has studied a modified form of the Ising model, where the system is divided
into cells within which the spin-spin interactions are constant. The shift of
the critical temperature as a function of the cell size is then calculated by
means of perturbation theory.  In three dimensions, the leading-order result of
Brout is recovered, namely a shift proportional to $1/q$. In the
next-to-leading term a logarithmic dependence on the coordination number is
obtained,
\begin{equation}
 T_c - T_c^{\rm MF} = \frac{a_1}{q} + a_2\frac{\ln q}{q^2} + \cdots \;,
\label{eq:th_shift_3d}
\end{equation}
whereas for $d=2$ the logarithm emerges already in the leading term,
\begin{equation}
 T_c - T_c^{\rm MF} = b_1\frac{\ln q}{q} + \cdots \;.
\label{eq:th_shift_2d}
\end{equation}
The latter expression is in perfect agreement with Eq.~(\ref{eq:shift_2d}),
whereas the logarithm in the higher-order term in~(\ref{eq:th_shift_3d}) has
not been found in Refs.~\cite{medran,thesis}. Since the logarithms in
Eqs.~(\ref{eq:th_shift_3d}) and~(\ref{eq:th_shift_2d}) apparently follow from
the same mechanism and the factor $\ln R$ in Eq.~(\ref{eq:shift_2d}) is
specific for the two-dimensional case (where all higher-order terms in the LGW
Hamiltonian are equally relevant), I conclude that there must be two different
sources for the logarithms, which happen to yield the same effect in $d=2$.
Indeed, the logarithms in Eqs.\ (\ref{eq:th_shift_3d})
and~(\ref{eq:th_shift_2d}) arise from counter terms canceling the infrared
divergences in the perturbation expansion. This appears to be intimately linked
to the infrared divergences occurring in massless super-renormalizable field
theories at rational dimensionalities~\cite{symanzik}. Actually, the treatment
of Ref.~\cite{medran} {\em does\/} account for logarithmic factors in $d=3$,
although at much higher order.  For systems with a large interaction range, the
first part of the renormalization trajectory passes close to the Gaussian fixed
point. Near this fixed point, only the $\phi^4$ term is relevant for $d=3$ and
all terms $\phi^n$ with $n>6$ are irrelevant. The marginal character of the
$\phi^6$ term produces a logarithmic range dependence in the shift of the
critical temperature.  However, since this logarithm stems from the term
quadratic in $\phi^6$ and the field $\phi$ is rescaled by a factor $R^{-1}$,
this contribution is extremely weak. An actual calculation shows that it leads
to a shift proportional to $\ln R/R^{18} \propto \ln q/q^6$. In addition, the
$\phi^6$ term will yield a correction of order $R^{-8}$. However, it may be
added that it is generally expected~\cite{bgz} that such high composite
operators have very little influence near the Ising fixed point.

Let me now briefly return to the leading shift $R^{-2d/(4-d)}$ obtained in
Ref.~\cite{medran}. It is instructive to note that this shift is consistent
with crossover arguments first given by Riedel and Wegner~\cite{riedel}.
Indeed, the Ginzburg criterion states that a crossover from classical to
nonclassical critical behavior occurs as a function of the crossover
parameter~$t^{(4-d)/2}R^d$. In terms of a more general formulation, this
parameter is written as $t^\phi/g$, with $\phi=(4-d)/2$ and $g=R^{-d}$. The
crossover exponent $\phi$ (not to be confused with the field~$\phi$),
introduced in Ref.~\cite{riedel}, is just the exponent $4-d$ of the relevant
operator driving the system away from the Gaussian fixed point (i.e., the
$\phi^4$ term in the LGW Hamiltonian), divided by the thermal exponent $y_t=2$.
Then, on general grounds~\cite{riedel,fisher-fss}, the shift of $T_c$ is
predicted to scale as $g^{1/\phi} \propto R^{-2d/(4-d)}$. This is another
indication that the shift terms in Eq.~(\ref{eq:th_shift_3d}) originate from a
different, complementary mechanism. In addition, it is noted that the
formulation in terms of the crossover exponent~$\phi$ can be carried even
further~(see, e.g., Ref.~\cite{cardy-book}). Indeed, for any thermodynamic
quantity~$P$ which is near the Ising critical point proportional to~$t^{x_{\rm
I}}$, the combined dependence on $g$ and~$t$ will be
\begin{equation}
 P \propto g^{(x_{\rm G}-x_{\rm I})/\phi} t^{x_{\rm I}} \;,
\end{equation}
where $x_{\rm G}$ denotes the $t$ dependence of~$P$ near the Gaussian fixed
point. In terms of $t$ and~$R$, this can be written as
\begin{equation}
 P \propto R^{d(x_{\rm I}-x_{\rm G})/\phi} t^{x_{\rm I}} \;,
\end{equation}
which yields, e.g., $m \propto R^{2d(\beta-1/2)/(4-d)}t^\beta$ and $\chi
\propto R^{2d(1-\gamma)/(4-d)} t^{-\gamma}$, recovering Eqs.\ 
(\ref{eq:mag-rangedep}) and~(\ref{eq:chi-rangedep}).

\section{Monte Carlo simulations}
\label{sec:mc}

\subsection{Simulational details}
I have carried out extensive simulations of 3D simple cubic lattices consisting
of $L \times L \times L$ lattice sites with periodic boundary conditions. Each
spins interacts equally with its $q$~neighbors lying within a distance $R_m$,
i.e., the system is described by the Hamiltonian~(\ref{eq:hamil}).  For the
simulations I have used the cluster MC algorithm introduced in
Ref.~\cite{ijmpc}. Its application to the present case is described in more
detail in the appendix of Ref.~\cite{medran}. In order to avoid lattice effects
I formulate the analysis in terms of an effective interaction range
$R$~\cite{mon},
\begin{eqnarray}
\label{eq:effrange}
R^2 &\equiv& \frac{\sum_{j \neq i} ({\bf r}_i - {\bf r}_j)^2 K_{ij}}%
                  {\sum_{j \neq i} K_{ij}} \nonumber \\
    &=& \frac{1}{q} \sum_{j \neq i} |{\bf r}_i - {\bf r}_j |^2
        \quad\quad \text{with } |{\bf r}_i - {\bf r}_j | \leq R_m \;.
\end{eqnarray}
It is easily seen that $\lim_{R \to \infty} R^2 = 3R_m^2/5$.
Table~\ref{tab:range} lists $R_m$, $q$, and $R$ for the first thirteen neighbor
shells which have been examined in the present work.

Several tests have been carried out to check the implementation of the
algorithm.  For $R_m^2=1$ exact results (for $L=3,4$) and accurate MC data are
given in Ref.~\cite{ic3d} and for $R_m^2=2,3$ alternative MC programs were
available, allowing the verification of the data for various system sizes. I
have carried out very long Monte Carlo simulations ($10^9$ and $10^8$ Wolff
clusters, respectively) for $L=4$ and $L=20$ for these ranges, at couplings
close to $K_c(R)$. On the other hand, if one takes into account all lattice
symmetries, an explicit summation over all states is feasible for $L=3$
($2^{27} \approx 1.34 \times 10^8$ configurations).  For this case, I have
carried out simulations for all ranges $1 \leq R_m^2 \leq 14$. No systematic
deviations could be observed.  The actual simulations were carried out for
systems up to $L=200$ (8~million spins); the number of samples was chosen
depending on the system size. As a rule of thumb, the amplitude ratio~$Q$ (to
be defined below) had a relative accuracy on the permille level for the largest
systems.

\subsection{Determination of the critical temperatures}
\label{sec:q-tc}
In order to analyze the range dependence of several quantities, an accurate
knowledge of the critical temperature for each single value of~$R_m$ is
required.  The critical temperatures of systems with interaction ranges
corresponding to the first thirteen neighbor shells have been determined using
the amplitude ratio $Q_L = \langle m^2 \rangle^2_L / \langle m^4 \rangle_L$.
For the 3D Ising universality class and a cubic geometry with periodic boundary
conditions, this quantity has, in the thermodynamic limit, the universal
critical-point value $Q = Q_{\rm I} = 0.6233~(4)$~\cite{ic3d}. As mentioned in
Sec.~\ref{sec:intro}, an accurate determination of the critical point is mainly
hampered by the requirement that one must reach the Ising limit, i.e., $L_{\rm
min} \approx R^4$. For the inner shells, the smallest system sizes that could
be used in the finite-size analysis were of the same order as in an analysis of
the 3D nearest-neighbor Ising model, i.e., $L \gtrsim 5$. For the remaining
shells, the smallest allowable system sizes, as determined from the quality of
the least-squares fits, followed the restriction $L \gtrsim R^4$ rather closely.
Only for the outermost shells this criterion could be slightly relaxed.  Thus,
the accuracy of the fit results decreases considerably with increasing
interaction range, because the finite-size data cover a much smaller range of
system sizes and all the accurate results for small system sizes have to be
excluded from the analysis. The least-squares fits were made using the
finite-size expansion for $Q$ given in Ref.~\cite{ic3d},
\begin{equation}
  Q_L(K)= Q + a_1 (K-K_c) L^{y_t} + a_2 (K-K_c)^2 L^{2y_t} + \cdots
            + b_1 L^{y_i} + b_2 L^{2y_i} + \cdots \;,
\label{eq:expan}
\end{equation}
where $K$ denotes the spin-spin coupling, $K_c$ the critical coupling,
and the $a_i$ and $b_i$ are nonuniversal (range-dependent) coefficients. The
exponents $y_t$ and $y_i$ are the thermal and leading irrelevant exponent,
respectively. They are approximately given by: $y_t = 1.587~(2)$ and $y_i =
-0.82~(6)$~\cite{ic3d}, where the latter exponent was kept fixed in all
analyses. Table~\ref{tab:qfit} shows my resulting estimates for $Q$ and $K_c$.
In the first place, one notes that all systems belong, within the statistical
accuracy, to the 3D Ising universality class. The critical couplings for the
first three shells are in agreement with the old series-expansion results of
Domb and Dalton. In order to improve the accuracy of the results, I have
repeated all analyses with $Q$ fixed.

\begin{figure}
\centering
\leavevmode
\epsfxsize\figurewidth
\epsfbox{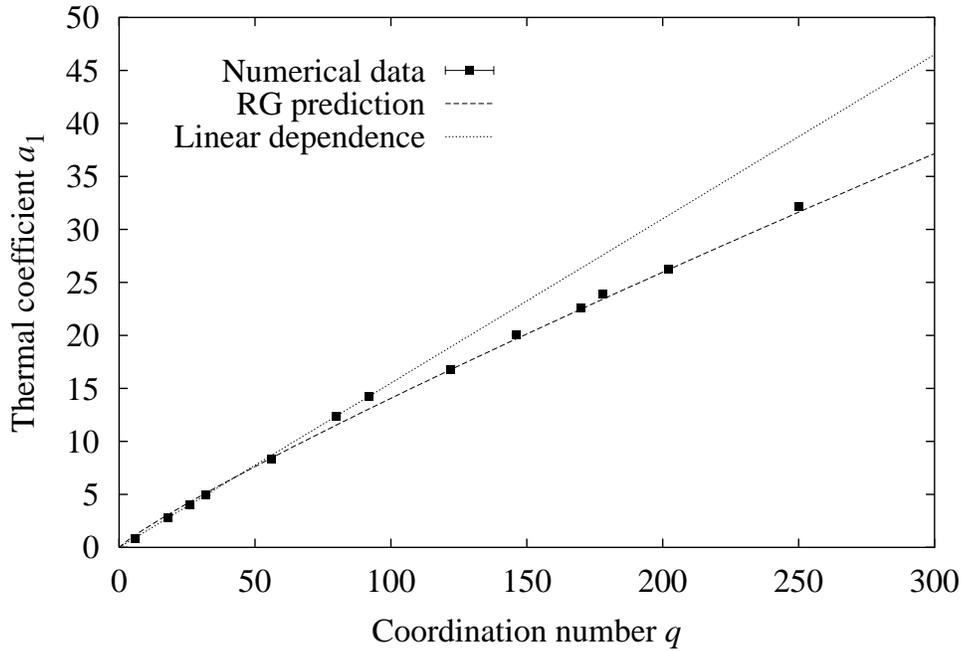}
\caption{The leading thermal coefficient in the finite-size expansion for the
amplitude ratio~$Q$, as a function of coordination number. The dashed curve
shows the RG prediction (valid in the large-$q$ limit) of
Ref.~\protect\cite{medran}. In order to appreciate the quality of this
prediction, a linear $q$ dependence is shown as well.}
\label{fig:tempcoef}
\end{figure}

The results of the finite-size analyses permit some additional tests of the
scaling predictions of Refs.~\cite{mon,medran}. Indeed, the range dependence of
the thermal coefficient $a_1$ in Eq.~(\ref{eq:expan}) should take the same form
as the first argument of the universal scaling functions~(\ref{eq:mag-fss})
and~(\ref{eq:chi-fss}). Upon expansion of such a scaling function one finds a
temperature-dependent argument of the form $a t L^{y_t} R^{-2(2y_t-d)/(4-d)}
\approx - a [(K-K_c)/K_c] L^{y_t} R^{-2(2y_t-d)/(4-d)}$, where $a$ is a
constant that does not depend on $R$.  Thus $a_1 = - a R^{-2(2y_t-d)/(4-d)} /
K_c \sim R^{-2(2y_t-d)/(4-d)} \times R^d \sim R^{2.652} \sim q^{0.884}$.
Figure~\ref{fig:tempcoef} shows $a_1$ as a function of the coordination
number~$q$. Both a curve $\sim q^{0.884}$ and a reference line with slope~1 are
shown; evidently the former describes the numerical data very well. Deviations
for relatively small~$q$ are not disturbing, since the RG predictions are only
valid in the limit of large interaction ranges and the small-$q$ data may also
exhibit some lattice effects.

\begin{figure}
\centering
\leavevmode
\epsfxsize\figurewidth
\epsfbox{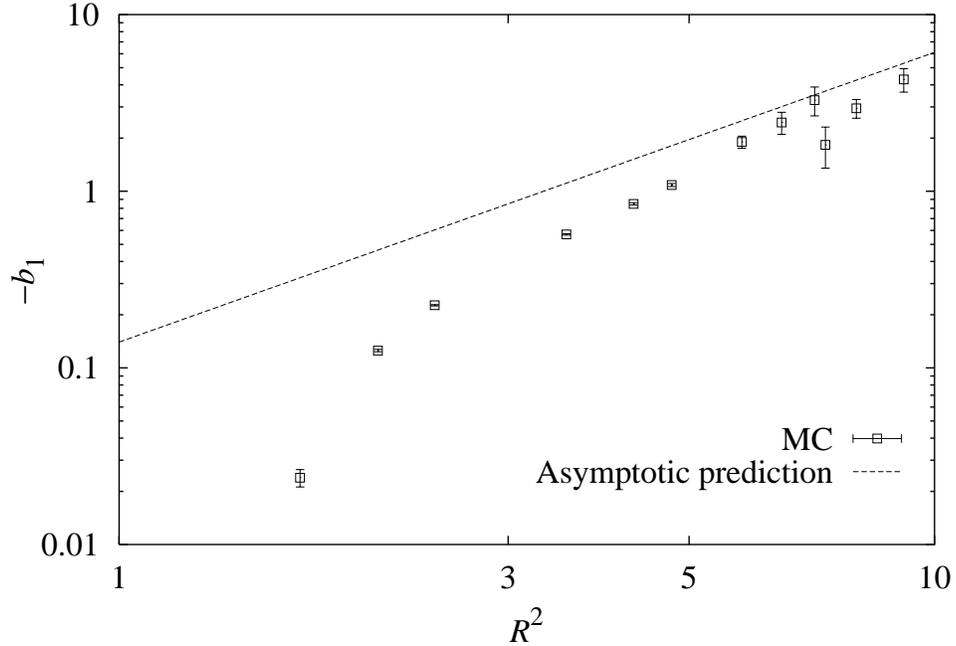}
\caption{Range dependence of the leading irrelevant field, cf.\ the second
argument in the right-hand side of Eq.~(\protect\ref{eq:mag-fss}). Note that
the result for $R=1$ is not shown, because it has the opposite sign. The dashed
line represents the asymptotic expression, $b_1 \propto R^{-4y_i/(4-d)}$, as
discussed in the text.}
\label{fig:asymcorr}
\end{figure}

Of particular interest is also the range-dependence of the coefficient~$b_1$ in
Eq.~(\ref{eq:expan}), because this coefficient is proportional to
$(u-u^*)/u^*$, where $u$ is the coefficient of the $\phi^4$ term in the LGW
Hamiltonian and $u^*$ is its fixed-point value~\cite{liu90}. As such, $b_1$
yields information on the $R$ dependence of the size and sign of the
corrections to scaling that appear in the Wegner expansion~\cite{wegner}. This
expansion describes the singular corrections to the asymptotic temperature
dependence of thermodynamic quantities close to the critical point.  For
example, if $u/u^* > 1$ the leading coefficient in the expansion for the
susceptibility will have a negative sign and hence the susceptibility
exponent~$\gamma$ will approach the Ising value from {\em above}, cf.\ also
Ref.~\cite{anisimov95}. On the other hand, if $u$ lies between the Gaussian and
the Ising fixed point, i.e., $0 < u/u^* < 1$, the sign of the first Wegner
correction will be positive and $\gamma$ will approach the Ising value from
below. In order to extract the $R$ dependence of $u$ from the coefficient
$b_1$, the RG scenario of Ref.~\cite{medran} has to be reconsidered.  It can be
shown that in the large-$R$ limit $u = u_0/R^4$. Because $u_0$ will exhibit a
remaining, weak $R$ dependence for small~$R$, I write it as $u_0(R)$. The first
part of the RG transformation is just a scale transformation in the
neighborhood of the Gaussian fixed point, which cancels the factor $1/R^4$
in~$u$. The $\phi^4$ coefficient can now be written as $u_0(R) = u^* + [u_0(R)
- u^*]$, which close to the Ising fixed point scales as $u_0 \to u_0' = u^* +
[u_0(R) - u^*] L^{y_i} R^{-4y_i/(4-d)}$~\cite{medran}. Thus, the coefficient
$b_1$ in Eq.~(\ref{eq:mag-fss}) is equal to $c [\bar{u}(R)-1]R^{-4y_i/(4-d)}$,
where $\bar{u}(R) \equiv u_0(R)/u^*$ and $c$ is a nonuniversal proportionality
constant.  For $R$ large, $\bar{u}(R)$ should go to a finite constant and hence
$b_1$ is expected to be proportional to $R^{-4y_i/(4-d)}$ in this limit. Just
as for most other quantities, it is difficult to accurately determine $b_1$ for
large interaction ranges, because the small system sizes have to be omitted
from the analysis.  Nevertheless, the results shown in Fig.~\ref{fig:asymcorr}
appear to be well compatible with the predicted $R$ dependence, with
$c[\bar{u}(\infty)-1] \approx -0.14$ (the latter estimate relies on the
assumption that the asymptotic limit has actually been reached for the largest
ranges shown in the figure).  Unfortunately, no estimate for $u_0(R)$ for
either $R=1$ (nearest-neighbor Ising model) or any other~$R$ is known to the
author, so that the overall constant~$c$ [which would have permitted the
calculation of $u_0(R)$ from $b_1(R)$] cannot be determined (cf.\ also
Ref.~\cite{shape3d}).

On the other hand, an estimate of the interaction range where $\bar{u}(R)=1$
does not depend on~$c$, and so it can be predicted with a reasonable accuracy
that this condition is fulfilled at $R^2 \approx 1.56$. The interest of this
point lies in the fact that the leading corrections to scaling should vanish
there, which in principle allows a much more accurate determination of critical
properties from numerical simulations.  This approach was used for the first
time in Ref.~\cite{ic3d}, where, amongst others, a spin-$\frac{1}{2}$ model
with nearest-neighbor coupling~$K_{\rm nn}$ and third-neighbor coupling~$K_{\rm
3n}$ was simulated.  The ratio $K_{\rm 3n}/K_{\rm nn}$ was set to 0.4, which in
hindsight proved to be somewhat too strong for fully suppressing the leading
corrections to scaling. A newer estimate yielded $K_{\rm 3n}/K_{\rm nn} =
0.25~(2)$ as an optimal choice~\cite{henk}. Further studies of these systems
were presented in Ref.~\cite{hasenb98}, where the coupling constant ratio was
systematically varied in order to eliminate the leading finite-size
corrections. This lead to an estimate of $K_{\rm 3n}/K_{\rm nn} \approx 0.27$.
Both estimates turn out to be in quite good agreement with my prediction for
{\em general\/} interaction profiles. Indeed, as follows from
Eq.~(\ref{eq:effrange}), an effective interaction range $R^2=1.56$ can be
obtained by, e.g., nearest-neighbor and next-nearest neighbor interactions with
$K_{\rm 2n}/K_{\rm nn} = 0.64$ or by nearest and third-neighbor interactions
with $K_{\rm 3n}/K_{\rm nn} = 0.29$.  This also explains the finding of
Ref.~\cite{ic3d} that $K_{\rm 2n}$ had to be chosen much larger than $K_{\rm
3n}$ to reach the same effect.

In this context it is of interest to review some series-expansion results for
the leading correction amplitudes for the magnetization, the susceptibility,
and the correlation length on simple cubic~(sc), body-centered cubic~(bcc), and
face-centered cubic~(fcc) lattices. Liu and Fisher~\cite{liu90} concluded that
the leading correction amplitudes are {\em negative\/} for the sc and bcc
lattices and gave various arguments that this also holds for the fcc lattice.
Furthermore, they argue that these amplitudes should vanish monotonically with
coordination number ($q=6,8,12$, respectively). This is indeed confirmed by the
fact that the data in Fig.~\ref{fig:asymcorr} {\em monotonically\/} approach
the predicted asymptotic $R$ dependence, apart from statistical scatter.
However, from the fact that for the sc lattice with $q=18$ ($R^2=5/3$) the
finite-size corrections have already changed sign, it would be expected that
the correction amplitudes for the fcc lattice are close to zero. In contrast,
both the results of George and Rehr~\cite{rehr} and Liu and
Fisher~\cite{liu89}, see Table~\ref{tab:rehr}, exhibit a relatively weak
variation with coordination number. On the basis of these results one would
certainly expect the leading corrections to vanish at much higher coordination
numbers. Thus, I conclude that, apart from the dependence on $q$ (or~$R$), the
value of $u$ has a rather strong dependence on the lattice structure as well.
For completeness, it may be remarked that the analyses of the Monte Carlo data
for the magnetization density and the susceptibility have revealed the same
monotonic $R$ dependence of the leading correction amplitude as that of the
quantity~$b_1$ discussed above.

\section{Range dependence at criticality}
\label{sec:rangedep}

\subsection{Critical temperature}
\label{sec:tc_shift}
The estimates for the critical coupling as given in Table~\ref{tab:qfit} can in
principle be used to verify the predictions for the shift of the critical
temperature. Because lattice effects are still relatively strong for the
interaction ranges studied here, the coordination number~$q$, appearing in,
e.g., Eq.~(\ref{eq:th_shift_3d}), cannot be used directly. It is expected that
these lattice effects disappear when the {\em effective\/} interaction
range~$R$ is used instead. Thus, the predicted shift is rewritten as:
\begin{equation}
 T_c^{-1} \equiv q K_c = 
 1 + \frac{c_0}{R^3} + \frac{c_1}{R^5} 
   + \frac{c_2 + c_3 \ln R}{R^6} + \cdots \;,
\label{eq:tc_shift}
\end{equation}
where I have used the inverse critical temperature to conform to the earlier
literature.  Unfortunately, it turns out that even in terms of~$R$ the
numerical data exhibit remarkably strong scatter for $R_m^2 \leq 10$, making it
impossible to obtain a sensible fit for the smaller interaction ranges. On the
other hand, for $R_m^2 > 10$, Eq.~(\ref{eq:tc_shift}) describes the data very
well. Because of the small variation of the $\ln R$ term over the fit range, it
was not possible to discern the coefficients $c_2$ and~$c_3$. Thus, I have
omitted $c_2$ altogether, which implies that this coefficient is absorbed into
an effective value of~$c_3$. The resulting fit yielded the values
$c_0=0.498~(2)$, $c_1=-5.7~(7)$, and $c_3=7.1~(9)$. Clearly, the last two
estimates suffer from the fact that (for the available values of~$R$) the last
two terms in Eq.~(\ref{eq:tc_shift}) lie quite close. Thus, it cannot be
excluded that the high values of $c_1$ and~$c_3$ are partially caused by a
mutual cancellation and that apart from the quoted statistical errors there is
a considerable systematic error.  Nevertheless, as will be seen below, the
accuracy of the resulting expression is sufficient to obtain rather precise
estimates for systems with larger interaction ranges.  In fact, if the results
for $R_m^2=9,10$ are also included in the least-squares fit and the lattice
effects are simply ignored, an essentially phenomenological interpolation
formula is obtained, which for larger ranges turns out to agree very well with
the first fit.

\begin{figure}
\centering
\leavevmode
\epsfxsize\figurewidth
\epsfbox{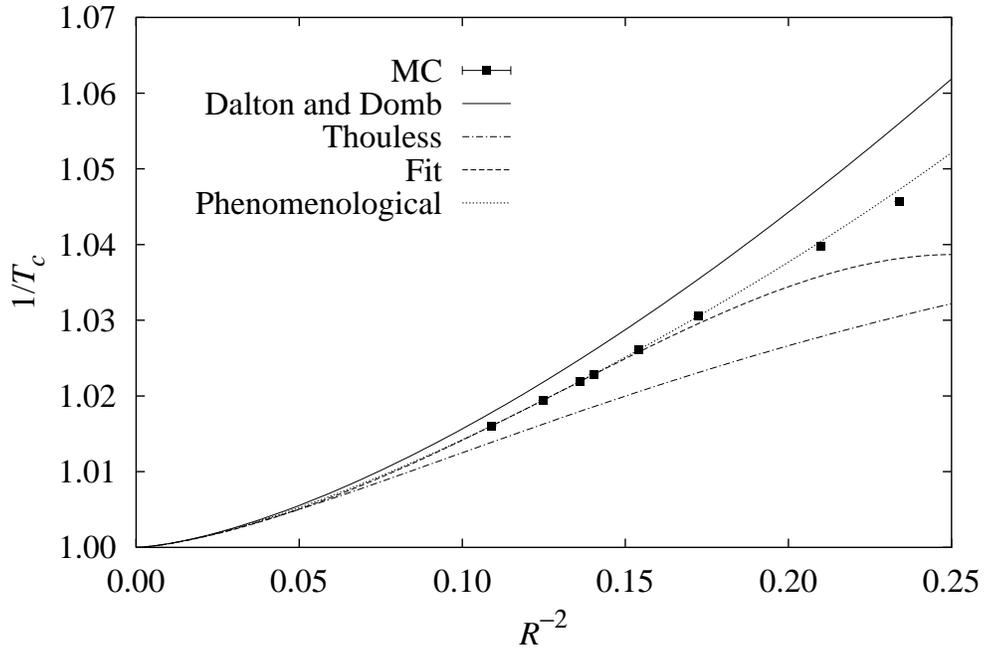}
\caption{Numerical results for the inverse critical temperature, normalized by
the mean-field critical temperature, as a function of the inverse squared
interaction range, together with the series-expansion results of Dalton and
Domb~\protect\cite{dalton66} and Thouless~\protect\cite{thouless}. The dashed
and the dotted lines indicate the results of the least-squares fits discussed
in Section~\protect\ref{sec:tc_shift}, where the dotted line is the
phenomenological description in which lattice effects have been ignored.}
\label{fig:tc_shift}
\end{figure}

In Refs.~\cite{dalton66,thouless}, series-expansion estimates are given for the
coefficients $c_0$ and $c_3$ in~(\ref{eq:tc_shift}). In terms of an expansion
in~$q$, Dalton and Domb found the value~$4.46$ for the leading coefficient
(confusingly, in later work~\cite{domb66,domb-dg3} the value~$3.5$ was quoted)
and for the prefactor of the logarithm Thouless obtained $-2000/27 \approx
-74.1$. To compare these values to $c_0$ and~$c_3$, I write $q+1 \approx
\frac{4}{3}\pi R_m^3 \approx \frac{4}{3}\pi (\frac{5}{3})^{3/2} R^3 \approx
9.013 R^3$. This yields $c_0 = 0.495$ and $c_3 = -2.74$.  In view of the
various approximations that have been made, the agreement for $c_0$ is truly
remarkable. Because of the above-mentioned cancellation effects and because of
the omission of $c_2$ in the fit, a sensible comparison for $c_3$ is not
possible. However, we note that also Thouless finds a relatively high value for
$c_3$.  Figure~\ref{fig:tc_shift} shows the various predictions for the shift
of the inverse critical temperature.

\subsection{Magnetization density}
\begin{figure}
\centering
\leavevmode
\epsfxsize\figurewidth
\epsfbox{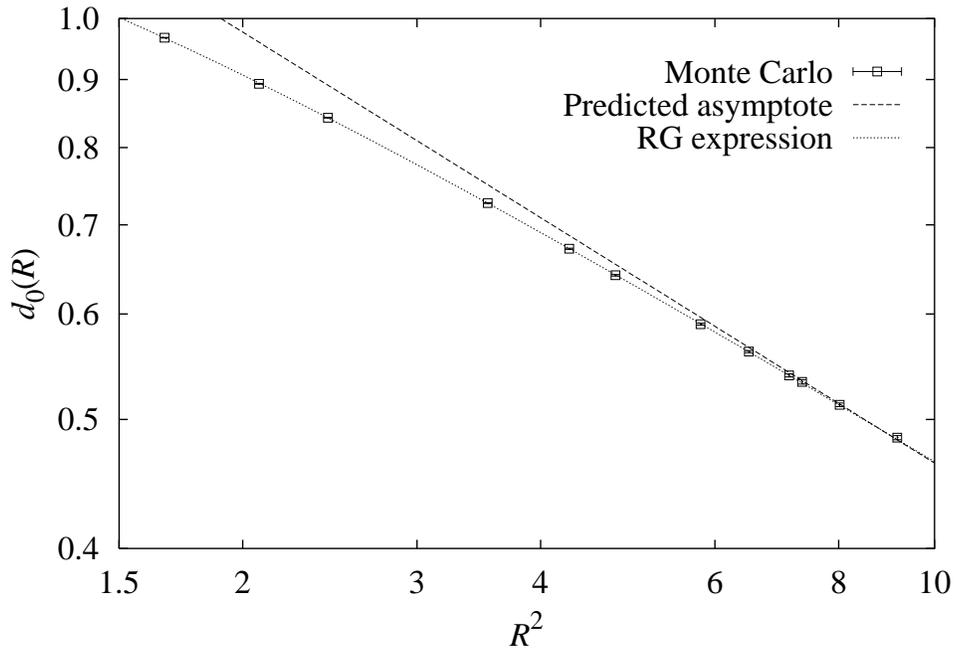}
\caption{Range dependence of the critical finite-size amplitude of the
magnetization density, together with the predicted asymptotic range dependence
(dashed line) and a fit of all the data points to the renormalization-group
prediction (dotted curve).}
\label{fig:mabs}
\end{figure}

In the Monte Carlo simulations, I have sampled the absolute magnetization
density $\langle |m| \rangle$. The dependence of this quantity on both $L$
and~$R$ is given by Eq.~(\ref{eq:mag-fss}), from which the following
finite-size expansion can be derived,
\begin{equation}
 m_L(K,R)= L^{y_h-d} \left\{ d_0(R) + d_1(R) [K-K_c(R)] L^{y_t} 
        + d_2(R) [K-K_c(R)]^2 L^{2y_t} + \cdots
        + e_1(R) L^{y_i} + \cdots \right\} \;.
\label{eq:mag-expan}
\end{equation}
For each single value of~$R$, I have fitted the numerical data to this
expression. The critical couplings obtained from this analysis are in agreement
with those shown in Table~\ref{tab:qfit}. The corresponding estimates for $y_h$
are listed in Table~\ref{tab:mfit}. The slight tendency of the estimates to
decrease with increasing~$R$, as well as the increasing uncertainties, can be
explained from the requirement that the smallest system size included in the
analysis must increase with~$R$. When the analyses were repeated with the
critical couplings fixed at the best values in Table~\ref{tab:qfit}, the
agreement of the estimates for $y_h$ (also shown in Table~\ref{tab:mfit}) with
the 3D~Ising value $y_h=2.4815~(15)$~\cite{ic3d} was even better. Thus, this
confirms the expectation that all these systems belong to the Ising
universality class. The critical amplitudes~$d_0(R)$ can be used to extract the
leading range dependence of the magnetization density. In order to maximize the
accuracy in these amplitudes, the results shown in Table~\ref{tab:mfit} were
obtained with the exponents $y_h$ and $y_t$ fixed at their Ising values (but
the critical coupling~$K_c$ was included as a free parameter). A fit of
$d_0(R)$ to the form $d_0(R) = d R^{x}$ for the largest three values of $R$
yielded $x=-0.87~(5)$, somewhat (although not significantly) smaller than the
predicted value $x=(3d-4y_h)/(4-d)= -0.926~(6)$. This shows that the asymptotic
regime, where higher-order corrections can be neglected, has not yet been
reached.  In general, the corrections are powers of $R^{-2}$~\cite{medran}:
\begin{equation}
\label{eq:rangecorr}
 d_0(R) = d R^x \left( 1 + \frac{A_1}{R^2} + \frac{A_2}{R^4} + \cdots \right)
\end{equation}
Expression~(\ref{eq:rangecorr}) with one correction term allowed me to obtain a
very acceptable fit ($\chi^2/{\rm DOF} \approx 0.6$) for {\em all\/} data
points with $2 \leq R_m^2 \leq 14$ and yielded $x=0.923~(5)$, in excellent
agreement with the RG prediction of Ref.~\cite{medran}. Figure~\ref{fig:mabs}
shows the MC results for $d_0(R)$ together with the asymptotic range dependence
and the full fit to the renormalization expression.

\subsection{Susceptibility}
\begin{figure}
\centering
\leavevmode
\epsfxsize\figurewidth
\epsfbox{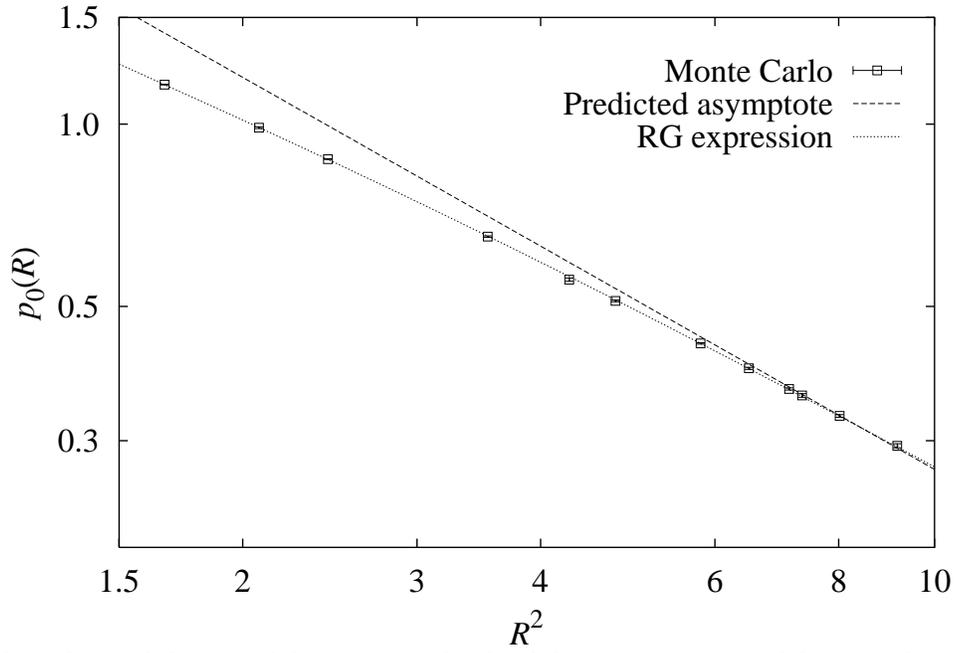}
\caption{Range dependence of the critical finite-size amplitude of the
magnetic susceptibility, together with the predicted asymptotic range
dependence (dashed line) and a fit of the data points to the
renormalization-group prediction (dotted curve).}
\label{fig:chi}
\end{figure}

At criticality, the magnetic susceptibility is directly proportional to the
average square magnetization. Thus, I have fitted the Monte Carlo data, for
each interaction range separately, to an expression of the form
\begin{equation}
 \chi_L(K,R)= s_0 + L^{2y_h-d} \left\{ p_0(R) + p_1(R) [K-K_c(R)]
              L^{y_t} + p_2(R) [K-K_c(R)]^2 L^{2y_t} + \cdots
              + q_1(R) L^{y_i} + \cdots \right\} \;,
\label{eq:chi-expan}
\end{equation}
where the additive constant $s_0$ originates from the analytic part of the free
energy. In the further analysis, this constant has been set to zero, because it
tends to interfere with the leading irrelevant term $q_1(R) L^{y_i}$. Just as
for the absolute magnetization density, I list estimates for the magnetic
exponent $y_h$ (Table~\ref{tab:chifit}). Although, as expected, the uncertainty
increases with~$R$, one observes that all estimates agree with the Ising value.
Also the critical couplings agree with those obtained from the fourth-order
amplitude ratio and $\langle |m| \rangle$. Thus, I have repeated all analyses
with $K_c$ fixed; the corresponding results for $y_h$ are shown in
Table~\ref{tab:chifit} as well. Finally, I have fixed the magnetic exponent at
$y_h=2.4815$ (but included $K_c$ as a free parameter) in order to obtain
accurate estimates for $p_0(R)$ (Table~\ref{tab:chifit}). Fitting a straight
line $pR^{-x}$ to the last three values yielded a slope~$-1.73~(9)$, which is
consistent with the prediction~$-1.852$ [Eq.~(\ref{eq:chi-fss})]. A fit formula
with one additional correction term, $pR^{-x}(1+ b R^{-2})$, allowed the
inclusion of several more data points and yielded $x=-1.92~(11)$. Both fits and
the numerical data are shown in Fig.~\ref{fig:chi}.

\subsection{Connected susceptibility}
\begin{figure}
\centering
\leavevmode
\epsfxsize\figurewidth
\epsfbox{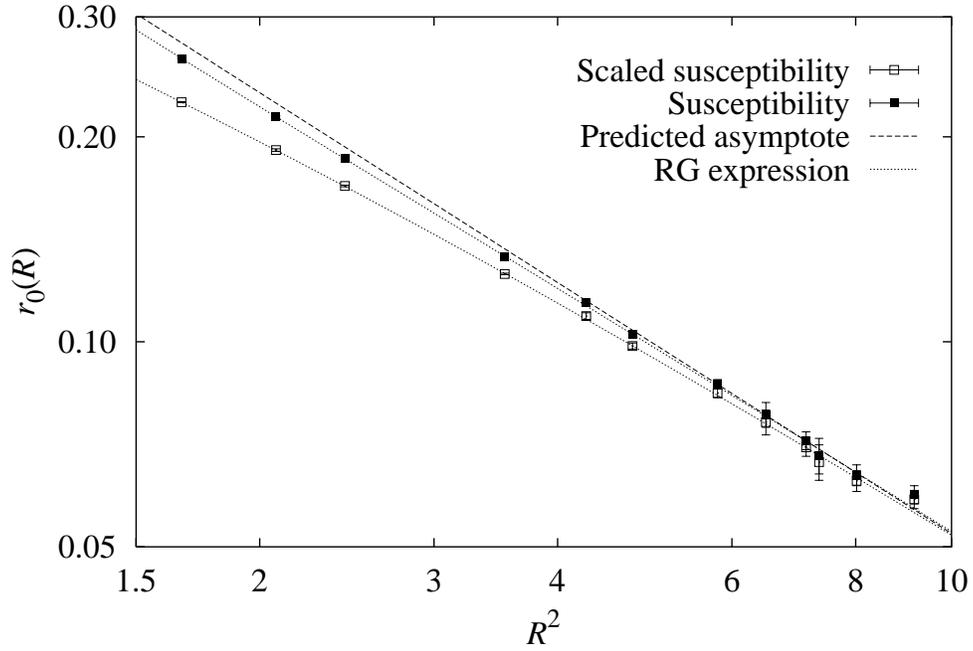}
\caption{Range dependence of the critical finite-size amplitude $r_0(R)$ of the
{\em connected\/} susceptibility, together with the predicted asymptotic range
dependence and a fit of the data points to the renormalization-group
prediction. Also the frequently-used {\em scaled\/} susceptibility is shown,
which clearly exhibits stronger deviations from the asymptotic range
dependence.}
\label{fig:con}
\end{figure}

In principle, the {\em connected\/} susceptibility, given by
\begin{equation}
 \tilde{\chi} =
 L^d \frac{\langle m^2\rangle - \langle |m| \rangle^2}{k_B T} \;,
\end{equation}
can be treated in the same way as the absolute magnetization density and the
susceptibility. The main drawback of this quantity, being the difference of two
fluctuating quantities, is that its statistical accuracy is relatively poor.
Nevertheless, the magnetic exponents extracted from the numerical data for the
individual interaction ranges are consistent with the Ising value and the
finite-size amplitudes can be used to determine the range dependence of the
connected susceptibility. As shown in Ref.~\cite{cross}, knowledge of this
dependence is very useful to determine the thermal crossover curve for the
susceptibility (which for $T<T_c$ is represented by~$\tilde{\chi}$) from data
for different~$R$, because it makes it possible to divide out the subleading
range dependence of this curve. Rather than giving the full details of the
analysis, I restrict myself to Fig.~\ref{fig:con}, which shows the critical
amplitudes together with the RG prediction fitted to it. Instead of
$\tilde{\chi}$, the so-called scaled susceptibility $k_B T \tilde{\chi}$ is
often considered.  It has been noted for the two-dimensional case~\cite{cross},
that the latter quantity exhibits considerably stronger deviations from the
asymptotic range dependence, which are caused by the shift of the critical
temperature.  Figure~\ref{fig:con} confirms that this also holds for the
three-dimensional case.

\section{Finite-size crossover}
\label{sec:cross}

\subsection{General considerations}
As stated in the introduction, the critical properties of the
equivalent-neighbor models obtained in the previous section can now be used to
find the finite-size crossover scaling functions describing the crossover from
a finite mean-field-like system to a finite Ising-like system at $T=T_c$. A
detailed description of this phenomenon has been given in Ref.~\cite{cross}.
Qualitatively this crossover can simply be understood from the observation that
systems with a linear size of the order of the interaction range are
essentially mean-field-like systems, which are turned into systems with a
short-range interaction if the system size grows beyond the interaction range.
RG considerations have shown that the crossover is ruled by a generalized
Ginzburg parameter $G \equiv LR^{-4/(4-d)}$, so that the mean-field regime
corresponds to $G \ll 1$ and the Ising regime to $G \gg 1$.  The expression
for~$G$ has also been obtained in Ref.~\cite{binder92}.  It is numerically not
feasible to observe the entire crossover regime in a system with fixed~$R$ by
merely varying the system size, since it spans several decades in the
parameter~$G$.  Thus, I construct the crossover curve by combining the results
for various interaction ranges, just as this has been done for the
two-dimensional case in Ref.~\cite{cross} and for the three-dimensional thermal
crossover in Ref.~\cite{chi3d}. Since it turns out that for $L \lesssim 20$ the
curves are affected by nonlinear finite-size effects, the smallest value of the
crossover parameter that can be reached with the interaction ranges studied in
the previous two sections is $20/(9.168)^2 \approx 0.24$. The true mean-field
regime, however, is only reached for {\em much\/} smaller~$G = {\cal
O}(10^{-4})$. Thus, I have carried out simulations for systems with effective
interaction ranges up to $R^2 = 323.81$ ($R_m^2=540$), corresponding to
coordination numbers as large as $q=52514$. Evidently, the Monte Carlo
algorithm introduced in Ref.~\cite{ijmpc} comes to its full glory here: The
simulation of three-dimensional systems with so many interactions present would
not have been feasible with either a Metropolis-type algorithm or a
conventional cluster-building algorithm. The actual crossover curves shown
below are obtained from a combination of the data for $2 \leq R_m^2 \leq 14$,
with system sizes between $L=20$ and~$L=200$, and additional data for 20
different interaction ranges $18 \leq R_m^2 \leq 540$. For the latter systems,
the critical coupling was determined using the extrapolation formula discussed
in Sec.~\ref{sec:tc_shift} and subsequently simulations were carried out for
$20 \leq L \leq 40$ at each single value of $R_m^2$.

An additional complication is formed by the regime $G \gg 1$. Whereas this part
of the crossover curve can easily be reached by simulating large system sizes
with very small interaction ranges, higher-order range dependences prevent the
direct use of these data for the construction of crossover curves. It was
recognized in Ref.~\cite{cross} that these are the same corrections that are
responsible for the deviations from the asymptotic range dependence in Figs.\ 
\ref{fig:mabs}, \ref{fig:chi}, and~\ref{fig:con}, so that this effect can be
removed by dividing the magnetization density by the factor in brackets in
Eq.~(\ref{eq:rangecorr}) and the other quantities by the corresponding
counterparts of this factor.

\subsection{Magnetization density}
\begin{figure}
\centering
\leavevmode
\epsfxsize\figurewidth
\epsfbox{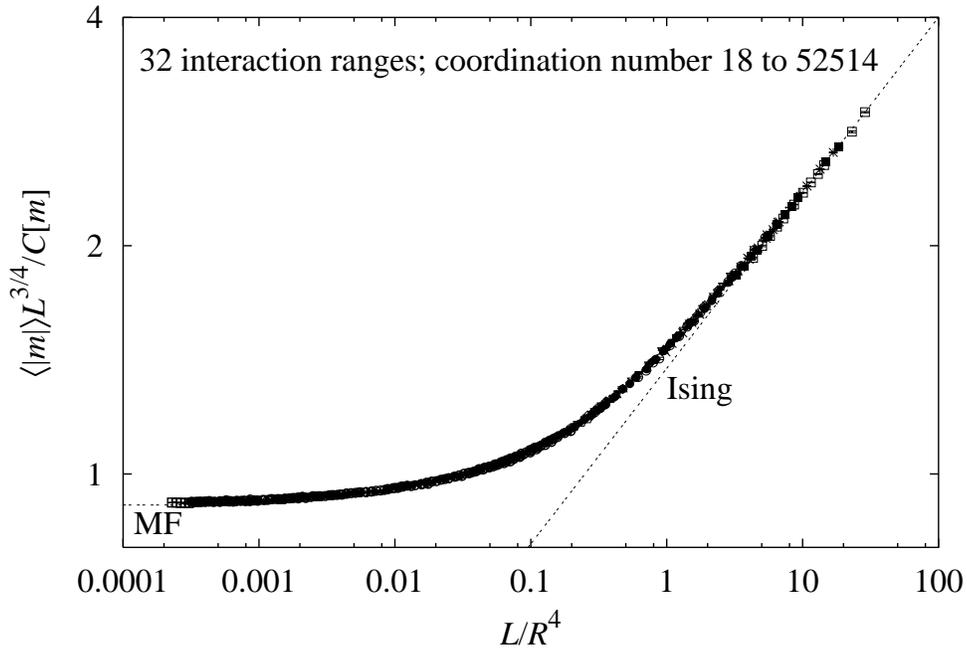}
\caption{Finite-size crossover curve for the absolute magnetization density
$\langle |m| \rangle$ multiplied by an appropriate power of the system size.
For very small interaction ranges (rightmost data points), higher-order range
dependences have been divided out, as indicated by the correction factor $C[m]$
(for a more extensive discussion of this topic the reader is referred to the
text). The crossover curve spans at least four decades in the parameter~$L/R^4$
and systems with a coordination number up to $q=52514$ had to be employed to
fully reach the mean-field limit.  The perfect collapse of all interaction
ranges and system sizes confirms the validity of the crossover description in
terms of a single parameter. The dashed lines denote the exact mean-field limit
(``MF'') and the Ising asymptote with slope $y_h-9/4$.}
\label{fig:mcross}
\end{figure}

As follows from Eq.~(\ref{eq:mag-fss}), the magnetization density $\langle |m|
\rangle$ at criticality is proportional to $L^{y_h-3}$ in the Ising regime. The
prefactor depends on the interaction range and scales as~$R^{9-4y_h}$. On the
other hand, $\langle |m| \rangle$ is independent of~$R$ in the mean-field
regime and just scales as $N^{-1/4} \propto L^{-3/4}$. If the crossover
behavior can indeed be described in terms of a single variable $G = L/R^4$, a
data collapse should be obtained for $\langle |m| \rangle L^{3/4}$. In the
mean-field regime, this quantity is independent of~$G$ and in the Ising regime
it scales as~$G^{y_h-9/4}$. The resulting curve for this quantity is shown in
Fig.~\ref{fig:mcross}. It is immediately clear that the data lie on a perfectly
smooth curve, confirming that the crossover is indeed ruled by the generalized
Ginzburg parameter~$G$. The correction parameter $C[m]=1+A_1 R^{-2}$ refers to
the higher-order range dependences which have been divided out, in order to
make the data for small~$R$ collapse on the same (Ising) asymptote. For large
interaction ranges this correction factor rapidly approaches unity. In the
graph I have included a line with slope $y_h - 9/4 = 0.2315$, indicating the
dependence on~$G$ in the Ising regime. Whereas no exact result exists for the
finite-size amplitude of this asymptote, it is possible to calculate its
counterpart in the mean-field regime, where it is found that~\cite{cross}
\begin{equation}
 \langle |m| \rangle L^{3/4} = 
  12^{1/4} \frac{\Gamma(\frac{1}{2})}{\Gamma(\frac{1}{4})}
  + {\cal O}(\frac{1}{L^{3/2}}) \;.
\end{equation}
Thus, $\langle |m| \rangle L^{3/4}$ should approach~$0.909\,891 \ldots$ in the
limit $G \to 0$. One indeed observes that the leftmost data points in the graph
lie already very close to this limit. Together with the collapse of all
numerical data onto a single curve, this also indicates that the simulations
for systems with large interaction ranges indeed have been carried out at the
correct temperatures; i.e., the extrapolation formula Eq.~(\ref{eq:tc_shift})
has yielded sufficiently accurate estimates for the critical temperatures for
$18 \leq R_m^2 \leq 540$. For the sake of clarity, it is stressed that for each
single value of~$R$, the simulations of the finite systems have been carried
out at the critical temperature of a system with that particular interaction
range in the thermodynamic limit.

\begin{figure}
\centering
\leavevmode
\epsfxsize\figurewidth
\epsfbox{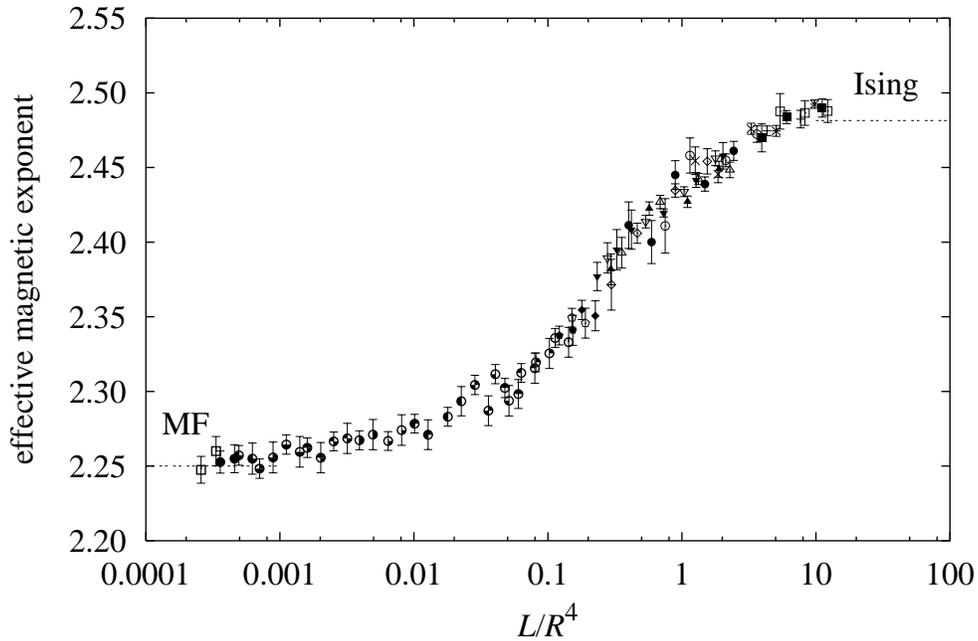}
\caption{The crossover behavior of the effective magnetic exponent as a
function of the finite-size crossover parameter.}
\label{fig:yhcross}
\end{figure}

As a side remark, I note that a much more sensitive description of the
crossover can be formulated in terms of so-called ``effective exponents''.
Originally introduced by Kouvel and Fisher~\cite{kouvel}, they have found
widespread use in experimental analyses (see, e.g., Ref.~\cite{anisimov95}) and
more recently also in the analysis of numerical results, cf.\ 
Refs.~\cite{chicross,cross,chi3d,shape3d}. Although these effective exponents
are usually defined in terms of the logarithmic derivative with respect to the
reduced temperature, an effective magnetic exponent can be introduced as
\begin{equation}
 y_h^{\rm eff} \equiv \frac{9}{4} +
 \frac{d \ln (\langle |m| \rangle L^{3/4})}{d \ln (L/R^4)} \;.
\end{equation}
In the mean-field regime, $y_h^{\rm eff}$ does {\em not\/} approach the
classical value~$y_h=1+d/2$, but the corresponding value $y_h^* = 3d/4$. This
directly related to the violation of hyperscaling in the mean-field regime and
can be explained from the dangerous-irrelevant-variable
mechanism~\cite{fisher-sa,bnpy,renorm}. This is clearly illustrated in
Fig.~\ref{fig:yhcross}, where a smooth interpolation between the value~$9/4$
and the Ising value~$2.4815$ is found.

\subsection{Susceptibility}
\begin{figure}
\centering
\leavevmode
\epsfxsize\figurewidth
\epsfbox{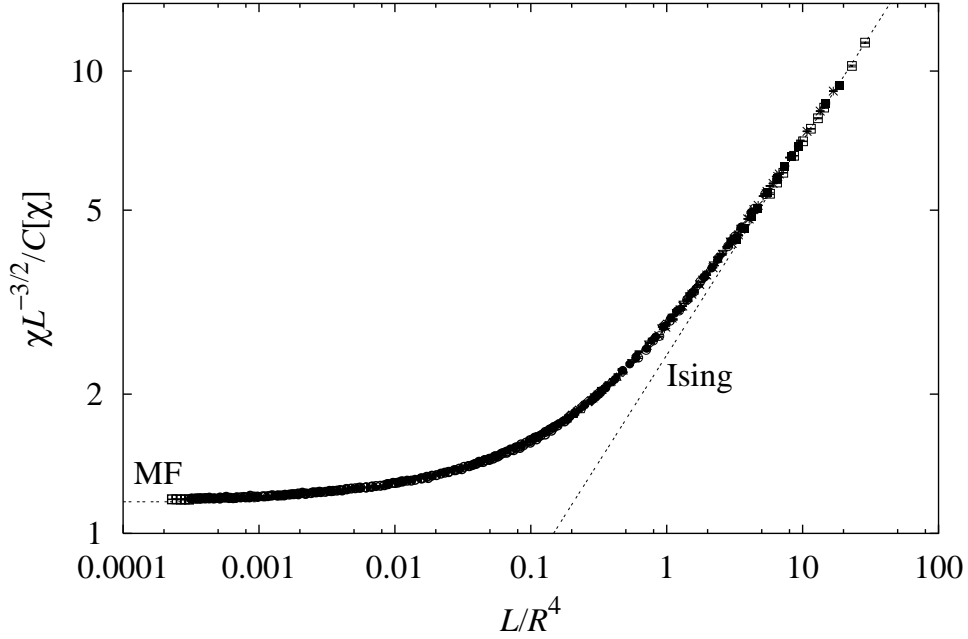}
\caption{Finite-size crossover curve for the magnetic susceptibility multiplied
by an appropriate power of the system size.  For very small interaction ranges
(rightmost data points), higher-order range dependences have been divided out,
as indicated by the correction factor $C[\chi]$. Just as in
Fig.~\protect\ref{fig:mcross}, systems with a coordination number up to
$q=52514$ had to be employed to fully reach the mean-field limit.  The perfect
collapse of all interaction ranges and system sizes confirms the validity of
the crossover description in terms of a single parameter. The dashed lines
denote the exact mean-field limit (``MF'') and the Ising asymptote with slope
$2y_h-9/2$.}
\label{fig:chicross}
\end{figure}

In a very similar way, the crossover function for the magnetic
susceptibility~$\chi$ at criticality can be obtained. Since it is proportional
to the average square magnetization density, it is independent of~$R$ in the
mean-field regime. In the Ising regime, it scales as $L^{2y_h-3}
R^{2(9-4y_h)}$, so that the quantity $\chi L^{-3/2}$ can be represented as a
function of the parameter~$G$. Indeed, upon application of the range-dependent
correction factor~$C[\chi]$, which has the same form as the factor between
brackets in Eq.~(\ref{eq:rangecorr}), a perfect data collapse is obtained, see
Fig.~\ref{fig:chicross}. The total crossover curve spans approximately four
decades in~$G$, just as for the magnetization density. The exact mean-field
result expected here is $\chi L^{-3/2} \to \sqrt{12} \Gamma(\frac{3}{4}) /
\Gamma(\frac{1}{4}) = 1.170\,829\ldots$, which is indeed well reproduced for
the data in the regime $G \to 0$. No nonlinear finite-size effects can be
observed, suggesting that these are (on the scale of the graph) negligibly
small for $L \geq 20$.

\subsection{Fourth-order amplitude ratio}
\begin{figure}
\centering
\leavevmode
\epsfxsize\figurewidth
\epsfbox{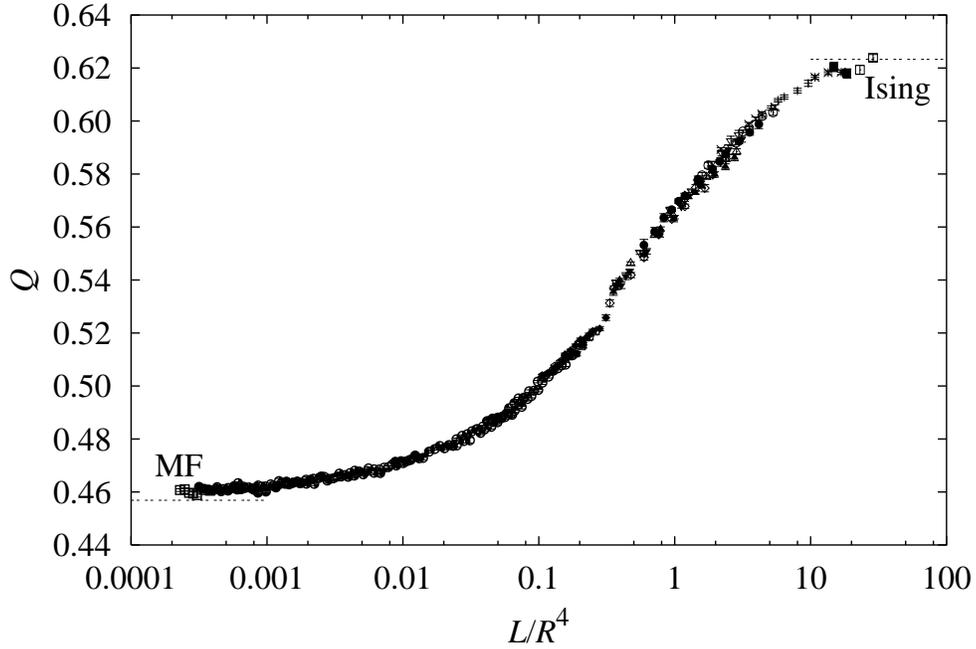}
\caption{Finite-size crossover curve for the amplitude ratio~$Q$. It
smoothly interpolates between the mean-field limit ($L/R^4 \ll 1$) and the
Ising limit ($L/R^4 \gg 1$).}
\label{fig:qcross}
\end{figure}

Rather than reproducing crossover curves for the connected susceptibility or
the spin-spin correlation function, which are very similar to those presented
in the previous two subsections, I prefer to pay some attention to the
crossover of the amplitude ratio~$Q$. This quantity, which is just a disguised
form of the fourth-order cumulant introduced by Binder~\cite{binder-q}, attains
trivial limiting values on either side of the critical temperature, but takes a
nontrivial universal value at criticality. Its Ising limit $Q_{\rm I} =
0.6233~(4)$ has already played an important r\^ole in Sec.~\ref{sec:q-tc},
where this parameter was used to determine the location of the critical point.
The critical value in the mean-field limit is known exactly, $Q_{\rm MF} =
0.456\,946\,58\ldots$~\cite{bzj,ijmpc}. Indeed, the full crossover from $Q_{\rm
MF}$ to $Q_{\rm I}$ as a function of $L/R^4$ can be observed, as illustrated in
Fig.~\ref{fig:qcross}. No correction term has been applied here, because it may
be expected that the correction terms for $\langle m^2\rangle^2$ and~$\langle
m^4 \rangle$ cancel each other to a large extent, cf.\ Fig.~8 in
Ref.~\cite{cross}. The less smooth appearance of the crossover curve compared
to that for the magnetization density and the susceptibility can mainly be
attributed to several other effects. Apart from the much larger scale of the
graph, it turns out that nonlinear finite-size effects are considerably
stronger for $Q$ than for other quantities. Further deviations are caused by
imperfections in the estimates for $T_c$ for large~$R$, which on this scale
become visible for the larger system sizes.

\section{Conclusions}
\label{sec:concl}
In this paper, I have presented a detailed determination of the critical
properties of the three-dimensional equivalent-neighbor model, which is a
generalization of the spin-$\frac{1}{2}$ Ising model, on a cubic lattice.
Monte Carlo simulations have been carried out for systems with up to thirteen
neighbor shells, corresponding to 250 equivalent neighbors.  All systems have
been shown to belong to the 3D Ising universality class.  An analysis of these
critical properties has yielded a coherent picture of their dependence on the
interaction range~$R$.  The shift of the critical temperature as a function of
interaction range, to which various mechanisms appear to contribute, has been
determined and compared to theoretical predictions. I have shown that the range
dependence of the critical finite-size amplitudes of the magnetization density
and the magnetic susceptibility conform very well to the theoretically expected
behavior. Also renormalization-group predictions for the variation of the
finite-size corrections with interaction range have been confirmed and an
estimate has been obtained for the effective interaction range at which the
leading finite-size corrections should vanish. The numerical results support
the expectation that the $\phi^4$~coefficient in the Landau-Ginzburg-Wilson
Hamiltonian varies monotonically with interaction range (or coordination
number) and scales for large ranges as~$1/R^4$.  Further Monte Carlo results
for systems with very large coordination numbers could be obtained by means of
an efficient simulation scheme.  These results enabled the mapping of the full
finite-size crossover curves for several quantities, including the magnetic
susceptibility and the fourth-order amplitude ratio. All these curves can be
described by a single crossover parameter~$L/R^4$ and interpolate smoothly
between mean-field and Ising-like behavior. Also the finite-size crossover
function for the effective magnetic exponent~$y_h$ has been obtained.

A very interesting and experimentally most relevant extension of the work
presented here is the case of {\em thermal\/} crossover, for which some first
results have appeared in Ref.~\cite{chi3d}. A more extensive analysis of this
case will be presented elsewhere~\cite{cross3d}. 

\acknowledgments 
It is a pleasure to acknowledge stimulating discussions with Kurt Binder and
Henk Bl\"ote. I wish to thank Andrea Pelissetto for illuminating correspondence
and John Rehr for sending me the series-expansion results of Ref.~\cite{rehr}
and for permission to publish them.  I thank the HLRZ J\"ulich for access to a
Cray-T3E on which the computations have been performed.

\newpage

\newpage
\begin{table}
\caption{The range of interaction $R_m$, the corresponding number of neighbors
 $q$, and the effective range of interaction $R$ for the thirteen neighbor
 shells examined in this work.}
\label{tab:range}
\renewcommand{\arraystretch}{1.2}
\begin{tabular}{r|r|r|c}
Shell & $R_m^2$ & $q$ & $R^2$                 \\ \hline
  1   &    1    &   6 & 1                     \\
  2   &    2    &  18 & $\frac{5}{3}$         \\
  3   &    3    &  26 & $\frac{27}{13}$       \\
  4   &    4    &  32 & $\frac{39}{16}$       \\
  5   &    5    &  56 & $\frac{99}{28}$       \\
  6   &    6    &  80 & $\frac{171}{40}$      \\
  7   &    8    &  92 & $\frac{219}{46}$      \\
  8   &    9    & 122 & $\frac{354}{61}$      \\
  9   &   10    & 146 & $\frac{474}{73}$      \\
 10   &   11    & 170 & $\frac{606}{85}$      \\
 11   &   12    & 178 & $\frac{654}{89}$      \\
 12   &   13    & 202 & $\frac{810}{101}$     \\
 13   &   14    & 250 & $\frac{1146}{125}$    \\
\end{tabular}
\end{table}

\begin{table}
\caption{The amplitude ratio $Q$ and critical coupling $K_c$ for the various
  ranges of interaction studied in this paper. The numbers in parentheses
  denote the errors in the last decimal places. The results for $R_m^2=1$ (3D
  nearest-neighbor Ising model) stem from Ref.~\protect\cite{ic3d}. The fourth
  column shows the estimates for $K_c$ obtained with $Q$ fixed at the value
  found in the same work (the error margins include the uncertainty in $Q$).
  For comparison, the estimates for $K_c$ given in
  Ref.~\protect\cite{domb66} are listed as well.}
\label{tab:qfit}
\begin{tabular}{r|d|d|d|d}
$R_m^2$ & $Q$         & $K_c$           & $K_c$           & $K_c$ \cite{domb66}
                                                                          \\ 
\hline
1       & 0.6233~(4)  & 0.2216546~(10)  &                 & 0.22171       \\
2       & 0.6238~(8)  & 0.0644223~(5)   & 0.0644220~(5)   & 0.06450       \\
3       & 0.6233~(8)  & 0.0430381~(4)   & 0.0430381~(4)   & 0.0432        \\
4       & 0.6224~(5)  & 0.03432668~(12) & 0.03432685~(15) &               \\
5       & 0.6216~(14) & 0.01892909~(7)  & 0.01892915~(4)  &               \\
6       & 0.621~(3)   & 0.01307105~(7)  & 0.01307111~(3)  &               \\
8       & 0.617~(4)   & 0.01130202~(8)  & 0.01130213~(3)  &               \\
9       & 0.608~(10)  & 0.00844691~(12) & 0.00844703~(4)  &               \\
10      & 0.614~(11)  & 0.00702798~(9)  & 0.00702798~(4)  &               \\
11      & 0.61~(2)    & 0.00601661~(14) & 0.00601663~(5)  &               \\
12      & 0.624~(11)  & 0.00574107~(7)  & 0.00574110~(4)  &               \\
13      & 0.618~(8)   & 0.00504666~(3)  & 0.00504666~(2)  &               \\
14      & 0.600~(14)  & 0.00406419~(4)  & 0.00406422~(2)  &               
\end{tabular}
\end{table}

\begin{table}
\caption{The leading correction amplitudes appearing in the Wegner expansion
for the magnetization ($T<T_c$, $a_m$), the magnetic susceptibility ($T>T_c$,
$a_\chi$) and the squared correlation length ($T>T_c$, $a_{\xi^2}$), for three
different lattice structures. The results for $a_m$ were taken from
Ref.~\protect\cite{liu89} and the results for $a_\chi$ and $a_{\xi^2}$ from
Ref.~\protect\cite{rehr}. The (slight) nonmonotonicity as a function of
coordination number in the latter two quantities, already noted in
Ref.~\protect\cite{liu90}, is probably not significant and also appears to
depend on the adopted choice for the susceptibility exponent~$\gamma$ (the
present results correspond to $\gamma=1.237$). The results for $a_m$ correspond
to the somewhat too high value $\beta=0.3305$, which can probably account for
the difference with the result $a_m \approx -0.203$ (for the sc lattice)
obtained in Ref.~\protect\cite{talapov}.}
\label{tab:rehr}
\begin{tabular}{lccc}
            & sc $(q=6)$ & bcc $(q=8)$ & fcc $(q=12)$ \\
\hline
$a_{m}$     & $-0.256$   & $-0.240$    & $-0.234$     \\
$a_{\chi}$  & $-0.108$   & $-0.119$    & $-0.114$     \\
$a_{\xi^2}$ & $-0.363$   & $-0.217$    & $-0.222$
\end{tabular}
\end{table}

\begin{table}
\caption{The magnetic exponent $y_h$ and the critical amplitude $d_0(R)$ of the
absolute magnetization density as a function of interaction range. The
estimates for $y_h$ in the third column have been obtained with $K_c$ fixed at
their best values given in Table~\protect\ref{tab:qfit}, whereas the critical
amplitudes have been obtained with $y_h$ fixed at its 3D Ising value.}
\label{tab:mfit}
\begin{tabular}{r|d|d|d}
$R_m^2$ & $y_h$      & $y_h$     & $d_0(R)$ \\
\hline
 2      & 2.479 (2)  & 2.479 (1) & 0.9674 (5)  \\
 3      & 2.479 (2)  & 2.481 (1) & 0.8933 (6)  \\
 4      & 2.475 (5)  & 2.479 (1) & 0.8424 (4)  \\
 5      & 2.477 (4)  & 2.480 (1) & 0.7269 (5)  \\
 6      & 2.476 (6)  & 2.483 (2) & 0.6716 (7)  \\
 8      & 2.472 (7)  & 2.484 (3) & 0.6415 (9)  \\
 9      & 2.46 (2)   & 2.480 (3) & 0.5895 (10) \\
 10     & 2.47 (2)   & 2.478 (3) & 0.5622 (10) \\
 11     & 2.47 (2)   & 2.471 (5) & 0.5395 (14) \\
 12     & 2.53 (4)   & 2.485 (6) & 0.5335 (20) \\
 13     & 2.47 (2)   & 2.480 (5) & 0.5128 (16) \\
 14     & 2.463 (15) & 2.475 (4) & 0.4845 (17)
\end{tabular}
\end{table}

\begin{table}
\caption{The magnetic exponent $y_h$ and the critical amplitude $p_0(R)$ of the
magnetic susceptibility as a function of interaction range. The estimates for
$y_h$ in the third column have been obtained with $K_c$ fixed at
their best values given in Table~\protect\ref{tab:qfit}, whereas the critical
amplitudes have been obtained with $y_h$ fixed at its 3D Ising value.
The data point for $R_m^2=1$ is taken from Ref.~\protect\cite{ic3d}.}
\label{tab:chifit}
\begin{tabular}{r|d|d|d}
$R_m^2$ & $y_h$      & $y_h$      & $p_0(R)$ \\
\hline
 1      &            &            & 1.5580 (15) \\ 
 2      & 2.479 (1)  & 2.479 (1)  & 1.1620 (7)  \\  
 3      & 2.481 (6)  & 2.484 (3)  & 0.9865 (32) \\  
 4      & 2.478 (6)  & 2.484 (2)  & 0.8752 (12) \\  
 5      & 2.481 (8)  & 2.481 (3)  & 0.6518 (18) \\  
 6      & 2.478 (13) & 2.478 (12) & 0.5534 (35) \\  
 8      & 2.484 (14) & 2.480 (2)  & 0.5105 (16) \\  
 9      & 2.46 (3)   & 2.476 (9)  & 0.4343 (12) \\  
 10     & 2.46 (2)   & 2.474 (4)  & 0.3951 (15) \\  
 11     & 2.46 (2)   & 2.47  (1)  & 0.3653 (16) \\  
 12     & 2.48 (2)   & 2.481 (6)  & 0.3564 (24) \\  
 13     & 2.46 (2)   & 2.484 (5)  & 0.3297 (16) \\    
 14     & 2.45 (4)   & 2.477 (6)  & 0.2943 (23)    
\end{tabular}
\end{table}

\end{document}